\documentclass[12pt,preprint]{aastex} 
\newcommand{\ergseccm}{erg\,s$^{-1}$\,cm$^{-2}$}
\newcommand{\etal}{et\,al.}
\newcommand{\halpha}{H$\alpha$}
\newcommand{\lsim}{\raise0.3ex\hbox{$<$}\kern-0.75em{\lower0.65ex\hbox{$\sim$}}}
\newcommand{\ujpbeam}{\,\,$\mu$Jy\,beam$^{-1}$}
\newcommand{\msun}{M$_{\odot}$}
\newcommand{\HII}{H~{\sc ii}}
\newcommand{\snu}{S$_{\nu}$}
\begin{document}
\slugcomment{ApJ Letters, in press}
\title{The Nature Of Radio Continuum Emission At Very Low Metallicity:
{\it VLA} Observations of I Zw 18} 

\author{John M. Cannon and Fabian Walter}
\affil{Max-Planck-Institut f{\"u}r Astronomie, K{\"o}nigstuhl 17, D-69117 
Heidelberg, Germany}
\email{cannon@mpia.de; walter@mpia.de}

\author{Evan D. Skillman}
\affil{Department of Astronomy, University of Minnesota,\\ 116 Church St. 
S.E., Minneapolis, MN 55455, USA}
\email{skillman@astro.umn.edu}

\author{Liese van Zee}
\affil{Department of Astronomy, Indiana University,\\ 727 East Third Street,
  Bloomington, IN 47405, USA}
\email{vanzee@astro.indiana.edu}

\begin{abstract}

We present the first resolved study of the radio continuum properties of 
I\,Zw\,18, the dwarf galaxy with the lowest known nebular metal abundance in 
the local universe.  New {\it Very Large Array} radio continuum images at 20 
and 3.6 cm are compared to various {\it Hubble Space Telescope} images, and we 
find a striking morphological similarity between high resolution \halpha\ and
short wavelength radio continuum emission, especially in the \halpha\ shell in 
the northwest region. We separate thermal and nonthermal components of the 
emission, and find a large synchrotron halo surrounding the galaxy. Comparison 
between \halpha\ and X-band fluxes suggests that the emission at 3.6\,cm is 
dominated by thermal processes; an additional synchrotron component dominates 
the flux at 20\,cm and produces a modest fraction of the detected flux at 
3.6\,cm.  The fluxes of three of the four major emission peaks show a mix of
thermal and nonthermal processes, while one shows a nearly flat spectral
index. The strong synchrotron component argues for active star formation 
throughout the disk for at least the last $\sim$ 30 Myr.  These sensitive 
observations provide a new, detailed view of the nature of radio continuum 
emission in the very low metallicity interstellar medium. Comparing with the 
literature, the role of metallicity in the evolution of radio continuum 
emission seems to be secondary to other factors such as recent star formation 
history and the presence or absence of outflows from star formation regions.

\end{abstract}						


\section{Introduction}
\label{S1}

With the lowest nebular metallicity known in the local universe [(O/H) $\sim$ 
0.02 (O/H)$_{\odot}$; {Skillman \& Kennicutt 1993}\nocite{skillman93}], 
I\,Zw\,18 plays a key role in the study of galaxy evolution.  Understanding 
the nature of the star-forming fragments predicted at high redshift requires a 
detailed study of nearby potentially ``young'' galaxies. Estimates of the age 
of the stellar population in I\,Zw\,18 based on single-star photometry suggest 
an age of less than 1 Gyr (e.g., {Hunter \& Thronson 1995}\nocite{hunter95}; 
{Aloisi \etal\ 1999}\nocite{aloisi99}; {{\" O}stlin 2000}; {Izotov \& Thuan 
2004}\nocite{izotov04b}).  

An alternative probe of the unobscured, recent (\lsim\ 30 Myr) star formation
in a galaxy can be obtained with radio continuum observations. Three types of 
radio continuum emission are seen in local star-forming galaxies (see {Condon 
1992}\nocite{condon92} for a detailed review).  First, thermal free-free 
emission (characterized by a nearly flat spectral index, $\alpha \sim -$0.1, 
where \snu\ $\sim\ \nu^{\alpha}$) is proportional to the total Lyman continuum 
flux and hence can be used as an accurate star formation rate estimator.  
Second, nonthermal synchrotron emission, produced in supernovae (SNe) 
explosions and remnants, shows a more negative spectral index ($-$1.2 $\lsim\ 
\alpha\ \lsim\ -$0.4).  Finally, free-free absorption can occur in young, 
dense, heavily-embedded clusters that show an inverted spectrum ($\alpha >$ 
0.0; {Turner \etal\ 1998}\nocite{turner98}, {Kobulnicky \& Johnson 
1999}\nocite{kobulnicky99c}, {Johnson \& Kobulnicky 2003}\nocite{johnson03b}).
Distinguishing between these emission mechanisms allows one to constrain the 
ages of the emitting components: inverted sources are very young (ages $\sim$ 
10$^5$ - 10$^6$ yr; {Johnson \etal\ 2001\nocite{johnson01}}); thermal regions 
have ages comparable to those of \HII\ regions ($\lsim$ 10 Myr); synchrotron 
radiation dominates in regions older than this.  

Considering the amount of observational attention that I\,Zw\,18 has received, 
it is somewhat surprising that no detailed investigation of its radio 
continuum properties exists in the literature.  There have been two previously 
published radio continuum detections, but both have been extracted from 
spectral line observations \citep{lequeux80,vanzee98a}.  In order to study the 
nature of radio continuum emission at the lowest available nebular 
metallicity, we have obtained sensitive {\it Very Large Array} ({\it 
VLA})\footnote{The National Radio Astronomy Observatory is a facility of the 
National Science Foundation operated under cooperative agreement by Associated 
Universities, Inc.} observations of I\,Zw\,18.  We compare these data to {\it 
Hubble Space Telescope} ({\it HST})\footnote{Based on observations with the 
NASA/ESA Hubble Space Telescope, obtained at the Space Telescope Science 
Institute, which is operated by the Association of Universities for Research 
in Astronomy, Inc. under NASA contract No. NAS5-26555.} images of the stellar 
and nebular emission.  Throughout this paper, we adopt a distance of 12.6 Mpc 
for I\,Zw\,18 \citep{ostlin00}; however, most of our results (spectral 
indices, comparison to {\it HST} imaging) do not require a tightly-constrained 
distance estimate.

\section{Observations and Data Reduction}
\label{S2}

\subsection{{\it VLA} Radio Continuum Data}
\label{S2.1}

{\it VLA} radio continuum imaging was obtained using the A, B, and C arrays on 
2003, August 3, and 2004, January 20 and April 20 (total on-source integration 
time $=$ 15.1 hours) for program AC681.  All reductions were performed using 
the Astronomical Image Processing System (AIPS) package. First, interference 
and bad data were removed.  Flux, gain and phase calibrations were then 
applied, derived from observations of 1331+305 (3c286; primary calibrator) and 
0921$+$622 (secondary calibrator). The L-band (20\,cm) and X-band (3.6\,cm) 
observations were obtained in two arrays, and these {\it u-v} databases were 
concatenated. The calibrated {\it u-v} data were then imaged and cleaned to 
produce matched-beam images at each frequency that balanced resolution and 
sensitivity. We analyze a set of ``high'' resolution (L-band: beam $=$ 
2.18\arcsec\,$\times$\,2.01\arcsec, rms $=$ 11.6\ujpbeam; X-band: beam $=$ 
2.17\arcsec\,$\times$\,2.07\arcsec, rms $=$ 7.4\ujpbeam) and a set of ``low''
resolution (beam $=$ 4.5\arcsec, L-band rms $=$ 19\ujpbeam, X-band rms $=$ 
12\ujpbeam) images.

\subsection{{\it HST} Imaging}
\label{S2.2}

We compare our radio observations to archival {\it HST} observations of 
I\,Zw\,18 from programs GO-5434 (P.I. Dufour), GO-6536 (P.I. Skillman), and 
GO-9400 (P.I. Thuan).  The detailed handling of the data from programs 6536 
and 5434 are presented in \citet{cannon02}.  We use the same narrowband 
\halpha\ image presented in that work for analysis here [total \halpha\ flux 
$=$ (3.26\,$\pm$\,0.3)\,$\times$\,10$^{-13}$ \ergseccm, implying a current 
star formation rate of $\sim$ 0.05 \msun\,yr$^{-1}$ \citep{kennicutt94}]. From 
program 9400, we use an {\it HST}/ACS F555W image for comparison with the 
stellar continuum.  Based on the positions of {\it HST} Guide Stars in the 
ACS field and the agreement of the radio and \halpha\ peaks, we conservatively
estimate the uncertainty in the coordinate solution to be better than 
0.5\arcsec\ rms. 

\section{The Nature of Radio Continuum Emission}
\label{S3}

\subsection{Global Flux Measurements}
\label{S3.1}

\begin{figure}
\epsscale{0.7}
\plotone{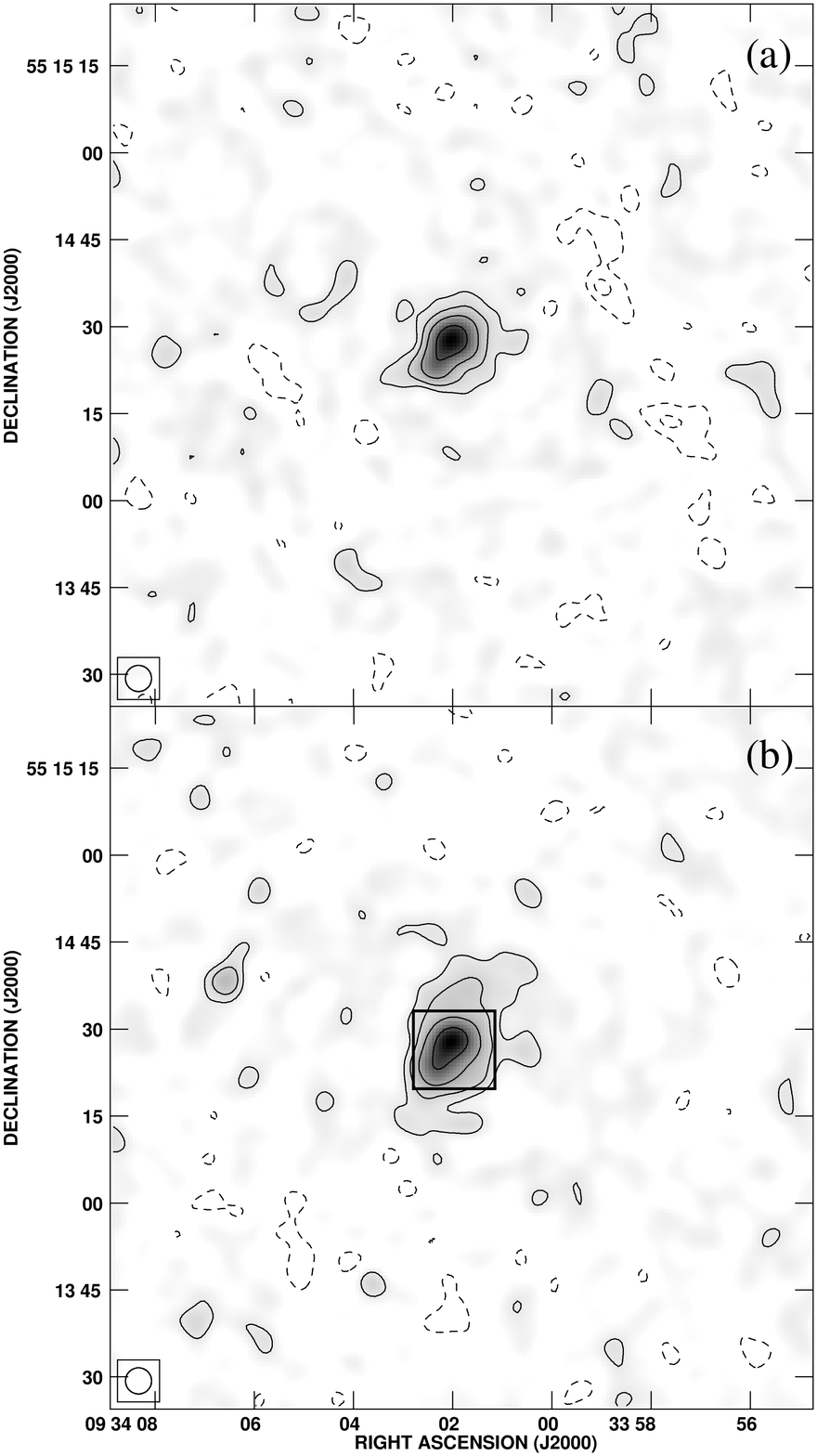}
\epsscale{1.0}
\caption{Low-resolution X-band (a) and L-Band (b) images of I\,Zw\,18,
 overlaid with contours at the ($-$4, $-$2, 2, 4, 8, 16)\,$\sigma$ levels 
(X-band 1\,$\sigma =$ 12\ujpbeam;  L-band 1\,$\sigma =$ 19\ujpbeam). The beam 
size is 4.5\arcsec\,$\times$\,4.5\arcsec\ and is shown at the bottom left. The
black box in (b) outlines the approximate field of view shown in
Figures~\ref{figcap2} and \ref{figcap3}. Note that the diffuse emission is
more spatially extended at 20\,cm than at 3.6\,cm.}
\label{figcap1}
\end{figure}

In Figure~\ref{figcap1} we present the low-resolution images overlaid with 
contours at various levels.  We measure global flux densities (by integrating 
continuous emission surrounding the galaxy after blanking at the 3\,$\sigma$ 
level) at L- and X-bands of 1.79\,$\pm$\,0.18 mJy and 0.78\,$\pm$\,0.08 mJy, 
respectively, corresponding to a global spectral index of $-$0.46\,$\pm$\,0.06
(see Table~\ref{t1}).  The continuum emission is more spatially extended at 
20\,cm than at 3.6\,cm (although the latter data are significantly more 
sensitive); indeed, emission above the 3\,$\sigma$ level is detected out to 
$\sim$ 12\arcsec\ (deconvolved) northwest of the emission peak at L-band 
(corresponding to a linear distance of $\sim$ 730 pc at the adopted distance),
but only to $\sim$ 7.5\arcsec\ at X-band.  As argued in the next section, this
is the result of a synchrotron halo surrounding the system. 

\placetable{t1}

\subsection{Thermal vs. Nonthermal Decomposition}
\label{S3.2}

Since the thermal radio continuum luminosity is proportional to the total 
photoionization rate, there exists a well-defined relation between radio 
continuum luminosity and other direct star formation rate indicators (e.g.,
\halpha\ emission).  In systems with little extinction in the optical, this 
relation can be used to estimate the expected thermal fraction of detected 
radio continuum emission.  In I\,Zw\,18, the low detected extinctions (A$_V$ 
\lsim\ 0.5 mag.; see {Cannon \etal\ 2002}\nocite{cannon02}) imply that 
\halpha\ provides a nearly complete census of the ongoing star formation.

We can estimate the expected thermal components at each frequency using
the relations presented in \citet{caplan86}. For a purely thermal radio 
continuum source in the absence of extinction, the \halpha\ flux and the flux 
density at a given frequency are related via:

\begin{equation}
\frac{j_{H\alpha}}{j_{\nu}} = 
\frac{(8.67\times10^{-9})(T)^{-0.44}}{(10.811 + 1.5\cdot ln(T) - ln(\nu))}
\label{eq1}
\end{equation}

\noindent where j$_{H\alpha}$ is the flux at \halpha\ in \ergseccm, j$_{\nu}$ 
is the flux density at the given frequency in Jy, T is the electron 
temperature in units of 10$^4$ K, and $\nu$ is the observed frequency in 
GHz. We adopt an intermediate electron temperature from \citet{skillman93} 
of 18500 K for the entire galaxy. Using the global \halpha\ flux measured from 
the {\it HST} image (see Table~\ref{t1}), we expect flux densities of 0.47 mJy 
at X-band and 0.56 mJy at L-band. Comparison with the measured X-band value of 
0.78 mJy suggests that more than half of the 3.6\,cm emission is thermal; 
conversely, only $\sim$ 30\% of the emission at L-band is expected to be 
thermal.

Given this expected value for thermal emission at 20\,cm, $>$ 70\% of the 
detected emission can be attributed to a synchrotron component. If this 
emission has a characteristic synchrotron spectral index ($\alpha = -$0.8), 
then this implies a synchrotron contribution to the detected flux at X-band 
of 0.29 mJy. This value can be compared with the difference between total and 
(predicted) thermal fluxes, 0.31 mJy; the agreement suggests that such a 
synchrotron component fits the data quite well.  

\subsection{Properties of Individual Radio Emission Peaks}
\label{S3.3}

\begin{figure}
\plotone{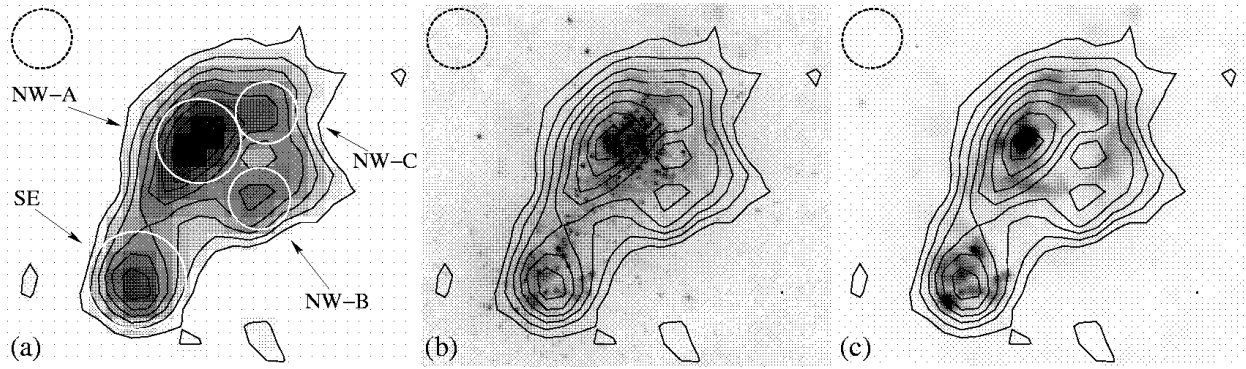}
\caption{High-resolution X-Band emission contours at the $-$3, 3, 4.5, 6, 7.5, 
9, 10.5, 12, \& 13.5\,$\sigma$ levels (1\,$\sigma =$ 7.4\ujpbeam), superposed 
on a grey-scale image of the emission (a), on an {\it HST}/ACS F555W (V) image 
(b), and on an {\it HST}/WFPC2 F656N (continuum-subtracted \halpha) image
(c). Labels and the white circles in (a) denote the apertures used to measure 
fluxes of individual peaks in the radio continuum and in \halpha; see 
\S~\ref{S3} and Table~\ref{t1}. The beam size is 
2.17\arcsec\,$\times$\,2.07\arcsec, and is shown at upper left.  The field of 
view (with north up and east to the left) is $\sim$ 0.9\,$\times$\,0.8 kpc at 
the adopted distance of 12.6 Mpc.}
\label{figcap2}
\end{figure}

\begin{figure}
\plotone{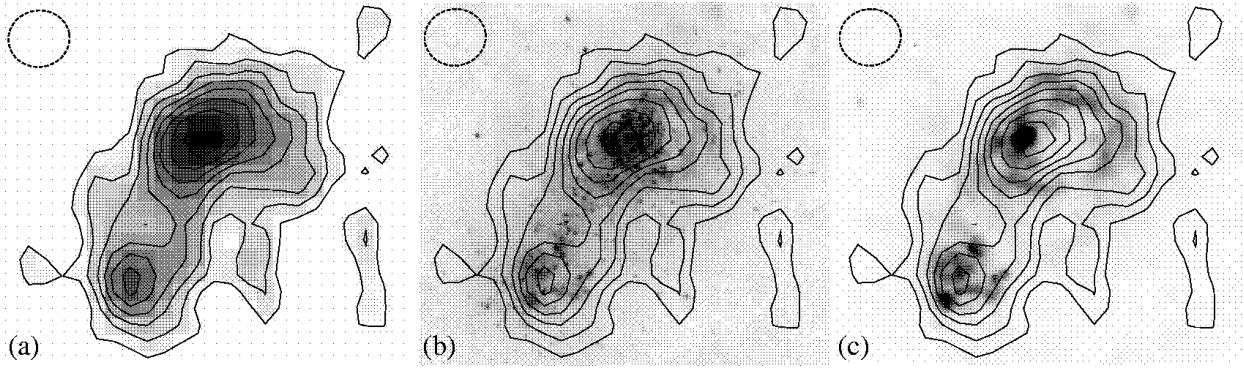}
\caption{Same as Figure~\ref{figcap2}, but for the L-band high-resolution
data (1\,$\sigma =$ 11.6\ujpbeam). Comparing to the X-band emission at similar 
resolution, it is clear that the individual emission peaks are less 
well-defined; we interpret this as the result of the additional synchrotron 
component at L-band.  The beam size is 2.18\arcsec\ $\times$ 2.04\arcsec, and 
is shown at upper left.}
\label{figcap3}
\end{figure}

We present higher resolution images of the radio continuum emission from 
I\,Zw\,18 in Figures~\ref{figcap2} and \ref{figcap3}, as well as overlays of 
the emission on {\it HST} \halpha\ and V-band images. We resolve the emission 
into 4 distinct peaks at this resolution. Examining these images, there is a 
striking correlation between \halpha\ and radio continuum emission at this 
sensitivity level; in all areas of high surface brightness \halpha\ emission, 
strong radio continuum emission is prevalent. Note also that the radio 
continuum emission closely follows the diffuse \halpha\ shell morphology in 
the northwest region.

The total flux densities of the individual peaks were measured using identical 
apertures the size of the beam or larger, centered on the peaks in the X-band 
image (see Figure~\ref{figcap2} and Table~\ref{t1}). The individual radio 
continuum peaks are found to have a mix of thermal and nonthermal spectral 
indices. Three of the sources (I\,Zw\,18 SE, NW-A, NW-C) have spectral indices 
$\alpha \sim -$0.27, implying a mix of thermal and nonthermal components.  
I\,Zw\,18 NW-B, however, is consistent with nearly purely thermal emission 
($\alpha = -$0.04\,$\pm$\,0.06).  

Sources I\,Zw\,18 SE and NW-A are clearly associated with collections of the 
high surface brightness \HII\ regions of the galaxy.  The high (aperture
integrated) \halpha\ equivalent widths found in \citet{cannon02} imply 
strong active star formation, while their nonthermal radio continua suggest 
that the areas are also rich in massive star feedback via SNe explosions and 
remnants.  This suggests that the star formation intensity was strong $\sim$ 
20-30 Myr ago, as well as at the present epoch.  This is in good agreement
with the results of resolved stellar population analysis, where recent 
investigations have found a star formation rate that was elevated during the 
last 10-100 Myr compared to that at present 
\citep{aloisi99,ostlin00,izotov04b}.  

\section{Discussion and Conclusions}
\label{S4}

The nature of radio continuum emission at very low metallicities is only 
beginning to be explored.  Comparing our results with those on the second-most
metal poor galaxy known, SBS\,\,0335$-$052 \citep{hunt04}, we find that both 
systems have a mix of nonthermal and thermal components, indicative of active 
massive star formation over the last $\sim$ 30 Myr.  However, the radio 
spectrum of SBS\,\,0335$-$052 has a thermal component that manifests itself in 
considerable absorption at L-band and a radio turnover between 1.4 and 4.8
GHz. While the present dataset has fewer frequencies, we would be sensitive to 
a similar spectral turnover in I\,Zw\,18 (as this would result in positive, 
rather than negative, spectral indices between 3.6 and 20\,cm).

The differences in the radio continuum properties of I\,Zw\,18 and 
SBS\,\,0335$-$052 are most easily explained as the results of the more
vigorous star formation event that is underway in the latter.  I\,Zw\,18 
appears to be undergoing star formation in discrete stellar clusters and \HII\ 
regions [see Table~\ref{t1} for \halpha\ luminosities of the radio emission 
peaks studied here, and {Cannon \etal\ (2002)}\nocite{cannon02} for those of 
other clusters], and infrared imaging \citep[e.g.,][]{ostlin00,ostlin05} does 
not suggest embedded regions that would be able to give rise to ``super star 
clusters'' (SSCs; see {O'Connell 2004}\nocite{oconnell04conf} for a recent 
review of SSC properties). SBS\,\,0335$-$052, on the other hand, hosts 
multiple SSCs \citep{thuan97a} and regions of exceptionally high extinction 
and embedded star formation for a very metal poor galaxy 
\citep{hunt01,plante02}.  Given that complex (and often inverted) spectral 
indices typically arise from such regions 
\citep{turner98,kobulnicky99c,johnson03b}, it is perhaps not surprising that 
the radio spectra of I\,Zw\,18 and SBS\,\,0335$-$052 appear to differ.  

To summarize, we have detected thermal and nonthermal emission in I\,Zw\,18, 
including a synchrotron halo.  This low-metallicity, low-dust system shows a 
radio continuum spectrum between 20 and 3.6\,cm that is typical of
star-forming dwarf galaxies; however, the radio properties differ markedly
from those found in the second most metal-poor galaxy, SBS\,\,0335$-$052.  
Although the sample is small, the variety of radio continua found at very low
metallicities suggests that there is not a strong metallicity dependence for 
the production of radio continuum emission. Indeed, it is not necessary that 
one should exist; thermal continua are produced by stars that have formed, 
while nonthermal continua are typically the result of the evolution of massive
stars. While there certainly exist differences in the way that the stars form 
at different metallicities, once the stars have formed, radio continuum 
emission appears to be ubiquitous (though complex and dependent on other 
factors) in massive star formation regions regardless of metallicity.  

\acknowledgements

The authors would like to thank Chip Kobulnicky for helpful discussions during
the planning stages of this project, and the anonymous referee for helpful 
comments. 

\clearpage


\clearpage

\clearpage

\clearpage

\clearpage
\begin{deluxetable}{lccccc}
\tabletypesize{\scriptsize}
\tablecaption{Properties of Radio Continuum Emission in I\,Zw\,18}
\tablewidth{0pt}
\tablehead{
\colhead{Parameter}         
&\colhead{I\,Zw\,18 SE\tablenotemark{a}}
&\colhead{I\,Zw\,18 NW-A\tablenotemark{a}} 
&\colhead{I\,Zw\,18 NW-B\tablenotemark{a}}
&\colhead{I\,Zw\,18 NW-C\tablenotemark{a}}
&\colhead{I\,Zw\,18 Total Galaxy}}
\startdata
\vspace{0.1 cm}
R.A. (J2000)\tablenotemark{a} &09:34:02.36 &09:34:02.10 &09:34:01.85 
&09:34:01.82 &09:34:02.0\\ 
Dec. (J2000)\tablenotemark{a} &$+$55:14:23.10 &$+$55:14:28.06 &$+$55:14:26.00 
&$+$55:14:29.06 &$+$55:14:28\\
Aperture Radius (\arcsec) &1.75 &1.5 &1.1 &1.1 &---\\ 
S$_{\rm L\,Band}$ (mJy) &0.17\,$\pm$\,0.02 &0.20\,$\pm$\,0.02 
&0.050\,$\pm$\,0.005 &0.079\,$\pm$\,0.008 &1.79\,$\pm$\,0.18\\
S$_{\rm X\,Band}$ (mJy) &0.11\,$\pm$\,0.01 &0.12\,$\pm$\,0.01 
&0.047\,$\pm$\,0.005 &0.050\,$\pm$\,0.005 &0.78\,$\pm$\,0.08\\
$\alpha$ (\snu\ $\sim\ \nu^{\alpha}$) &$-$0.27\,$\pm$\,0.06 
&$-$0.29\,$\pm$\,0.06 &$-$0.04\,$\pm$\,0.06 &$-$0.26\,$\pm$\,0.06 
&$-$0.46\,$\pm$\,0.06\\ 
\halpha\ Flux (10$^{-14}$ \ergseccm) &7.7\,$\pm$\,0.8 
&7.8\,$\pm$\,0.8 &1.3\,$\pm$\,0.2 &2.9\,$\pm$\,0.3 &32.6\,$\pm$\,3.3\\
\enddata
\label{t1}\vspace{-0.2 cm}
\tablenotetext{a}{See Figure~\ref{figcap1} for positions.  The central
position of radio continuum emission peaks are derived from Gaussian fitting.
The coordinates of the I\,Zw\,18 system are taken from the NASA Extragalactic 
Database.}\\ 
\end{deluxetable}
\end{document}